# Enzyme-Based Logic Analysis of Biomarkers at Physiological Concentrations: AND Gate with Double-Sigmoid "Filter" Response

Jan Halámek,[a] Oleksandr Zavalov,[b] Lenka Halámková,[a,c]
Sevim Korkmaz,[a] Vladimir Privman,[b] Evgeny Katz[a]

[a] Department of Chemistry and Biomolecular Science,

[b] Department of Physics, and

[c] Department of Biology, Clarkson University, Potsdam, NY 13699, USA

## Abstract

We report the first realization of a biomolecular **AND** gate function with double-sigmoid response (sigmoid in both inputs). Two enzyme biomarker inputs activate the gate output signal which can then be used as indicating liver injury, but only when *both* of these inputs have elevated pathophysiological concentrations, effectively corresponding to logic-**1** of the binary gate functioning. At lower, normal physiological concentrations, defined as logic-**0** inputs, the liver-injury output levels are not obtained. High-quality gate functioning in handling of various sources of noise, on time scales of relevance to potential applications is enabled by utilizing "filtering" effected by a simple added biocatalytic process. The resulting gate response is sigmoid in both inputs when proper system parameters are chosen, and the gate properties are theoretically analyzed within a model devised to evaluate its noise-handling properties.

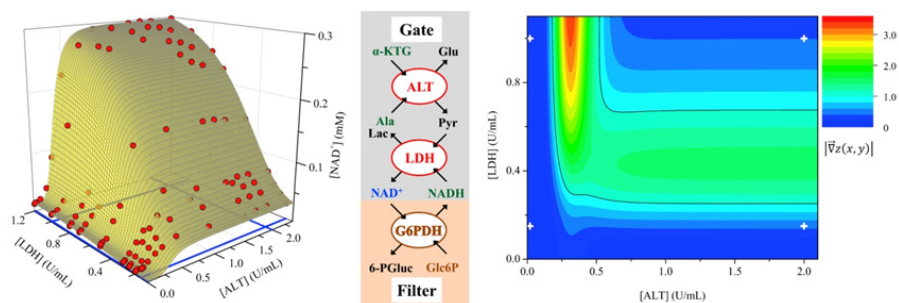

**INTRODUCTION**

Molecular[1] and biomolecular[2] information processing systems have recently received significant attention as promising research avenues in unconventional computing.[3,4] Chemical[5-9] and biochemical[10-14] systems realizing binary logic gates and networks have been investigated for novel computational and signal processing designs. Biomolecular systems have included those based on proteins/enzymes,[14,15] DNA,[13,16] RNA[17] and even whole cells.[18,19] Biomolecular computing systems in biochemical and biotechnological environments[20] promise design of novel biosensors capable of multiplexing and processing several biochemical signals in the binary format, **0** and **1**, with the information processing carried out by (bio)chemical processes rather than electronics.[21,22] Specifically, biomolecular logic has been explored[23-31] for prospective biomedical/diagnostic applications aiming at analysis of biomarkers characteristic of various pathophysiological conditions of interest in diagnostics of diseases or injuries.

Most model systems presently considered for (bio)chemical computing have logic-**0** values of chemical concentrations selected as the absence of reacting species, and logic-**1** defined as experimentally convenient nonzero concentrations of reactants. Systems designed for logic analysis of biomedical conditions should have logic-**0** and **1** concentrations (or possibly ranges of concentrations) correspond instead to normal physiological and pathophysiological conditions, respectively. In certain biosensor systems, e.g., in the pregnancy tests, the "digital" diagnostic YES/NO is easily achieved because the biomarker concentrations are well-resolved. However, in most cases the difference between **0** and **1** input ranges is not large as compared to the levels of noise, making the differentiation of the YES/NO answers difficult, unless careful optimization is performed.[31]

Optimization of biochemical reaction processes for digital sensing can be achieved in several ways. System parameters can be selected appropriately. For instance, the readout time can be adjusted to measure the signals when they are substantially different due to different kinetics of processes involved.[26,28] However, the problem of the output signal discrimination can be resolved more efficiently by adding "chemical-filter" reaction steps which modify convex response functions characteristic of most (bio)catalytic processes, to sigmoidal.[32] These novel (bio)chemical "filter" systems have recently been designed[33-36] and optimized as standalone elements for inclusion in biochemical logic networks. Specifically, we have demonstrated that integration of a filter system with digital biosensing approach can significantly improve performance, enabling differentiation between output-**0** and **1** values corresponding to normal physiological and pathophysiological concentrations of biomarkers[36] for liver, as well as abdominal trauma and soft-tissue injuries.

Two appropriate biomarkers jointly provide a better indication for each of these injuries, and can be inputted into the **AND** gate function for a YES/NO determination of the presence of that specific condition based on the gate's output. Specifically, in the present work we focus on the best-studied case in this context,[36] that of a pair of liver-injury biomarkers,[37,38] enzymes alanine transaminase (ALT) and lactate dehydrogenase (LDH). Elevated levels of both jointly have been used to indicate liver injury.[39,40] We consider a model system[28] in aqueous solution rather than in serum.[26] The aqueous system has not only been realized as the **AND** gate utilized to detect the presence of both enzymes (here, by optical means), but has also been coupled to a signal-responsive polymer-brush thin-film deposited on an electrode.[41] The thin-film has then acted as a switchable interface electronically amplifying the chemical changes generated by the biochemical logic gate. This particular application requires significant changes in the **AND**-gate chemical product in order to make the thin-film permeable for redox species. As a result, large gate times are appropriate, for which, however, the precision of the **AND** function realization was found not satisfactory. The added "filter" process then not only improved but for these gate times actually *enabled* a precise **AND** binary logic realization.[36]

Here we report a detailed mapping of the response function of this **AND** gate for the physiologically relevant input levels. We demonstrate that the added filter process makes it *double-sigmoid* (sigmoid in both inputs), which, to our best knowledge, is the first such demonstration in the literature for a set of few coupled biocatalytic reactions. We develop a kinetic model specifically designed to study the binary-logic gate-response properties of the system, and we use it to fit the experimental data to quantify the inherent noise, the input-to-output noise amplification,[32,42] and the tolerance properties of the realized **AND** gate as a component for biochemical logic.

**EXPERIMENTAL**

*Chemicals and reagents used:* Alanine transaminase (ALT) from porcine heart (E.C. 2.6.1.2), lactate dehydrogenase (LDH) from porcine heart (E.C. 1.1.1.27), glucose-6-phosphate dehydrogenase (G6PDH) from *Leuconostoc mesenteroides* (E.C. 1.1.1.49), β-nicotinamide adenine dinucleotide reduced dipotassium salt (NADH), L-alanine (Ala), D-glucose-6-phosphate (Glc6P), α-ketoglutaric acid (α-KTG), *tris*(hydroxymethyl)-aminomethane (Tris-buffer) and other standard inorganic/organic reactants were purchased from Sigma-Aldrich and used as supplied. Ultrapure water (18.2 MΩ·cm) from NANOpure Diamond (Barnstead) source was used in all of the experiments.

*Instrumentation and measurements:* A Shimadzu UV-2450 UV-Vis spectrophotometer with a TCC-240A temperature-controlled holder and 1 mL poly(methyl methacrylate) (PMMA) cuvettes were used for all optical measurements. All the optical measurements were performed at $37.0 \pm 0.2$°C mimicking physiological conditions and all reagents were incubated at this temperature prior to measurements.

*Composition and operation of system:* Scheme 1 (see page 22) shows the sequence of biocatalytic processes involved in the **AND** gate function, catalyzed by the two input enzymes, and the added filter process catalyzed by an additional enzyme, glucose-6-phosphate dehydrogenase (G6PDH). Further details of the system functioning[36] are presented below.

The system was realized in Tris-buffer, 100 mM, pH 7.4, used also as the reference background solution for our optical measurements. The "gate machinery" reactants were dissolved in the solution: Ala (200 mM), α-KTG (10 mM), and NADH (0.3 mM), in addition to the input enzymes. The filter-process chemicals when added, were G6PDH (20 U·mL$^{-1}$) and Glc6P (1.2 mM). The logic **0** and **1** levels of the input enzymes added were, respectively, ALT: 0.02 and 2 U·mL$^{-1}$, and LDH: 0.15 and 1 U·mL$^{-1}$. These were selected to correspond to meaningful circulating levels of these biomarkers under normal physiological and pathophysiological conditions, respectively.[43-45] For mapping out the gate-function properties, varying concentrations of both inputs were actually studied: see below. The output signal, [NAD$^+$], was quantified by detecting the decrease in the concentration of NADH, measured optically at λ = 340 nm. It was calculated by using the extinction coefficient,[46] 6.22 mM$^{-1}$cm$^{-1}$, for NADH.

In order to map the response-surface of the biocatalytic cascade out, concentrations of both biomarker-inputs (ALT and LDH) were varied starting at or somewhat below the logic-**0** and increasing somewhat above the logic-**1** values. For ALT the following concentrations were used: 0.01, 0.02, 0.04, 0.1, 0.2, 0.3, 0.5, 1.0, 1.6, 1.8, 2.0 and 2.4 U·mL$^{-1}$. For LDH the following concentrations were used: 0.15, 0.2, 0.25, 0.3, 0.5, 0.7, 0.8, 1.0 and 1.2 U·mL$^{-1}$. The experiments were performed with all the combinations of these input concentrations, thus resulting in the data array of 12×9 time-dependent experimental sets, shown in Figures 1 and 2 for the gate time (see pages 23-26 for the figures).

**MODEL SYSTEM FOR LIVER INJURY DETECTION**

We consider a biocatalytic cascade to detect the elevated levels of the two enzyme inputs, ALT and LDH. Their simultaneous increase in concentration, from normal to pathophysiological levels provides[39,40] evidence of liver injury. The sequence of the biochemical processes is shown in the "Gate" section of Scheme 1:

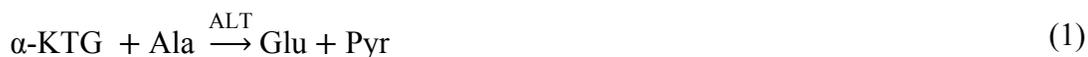
$$\alpha\text{-KTG} + \text{Ala} \xrightarrow{\text{ALT}} \text{Glu} + \text{Pyr} \tag{1}$$

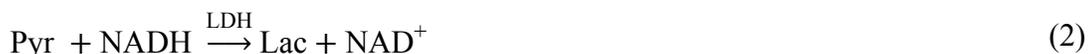
$$\text{Pyr} + \text{NADH} \xrightarrow{\text{LDH}} \text{Lac} + \text{NAD}^+ \tag{2}$$

Oxidation of NADH is followed by measuring the change in the absorbance, yielding the concentration of the output, NAD$^+$, which is produced only in the presence of the two input enzymes (and the "gate machinery" reactants α-KTG, Ala, NADH). However, here the logic-**0** values of the inputs are not the absence of the enzymes, but rather their presence at non-elevated physiological levels. Therefore, NAD$^+$ is produced not only for the **1,1** combination of the input signals, but also, at smaller quantities, when the inputs are supplied at the **0,0**, **0,1** and **1,0** combinations.

The response function for inputs varying from somewhat below their logic-**0** levels to somewhat above logic-**1**, has been mapped out experimentally as shown in Figure 1, for system parameters selection explained later. While data were collected for varying times, Figure 1 illustrates the result for the "gate time" of $t_g$ = 600 sec, relevant for applications involving signal-responsive membranes.[41] It is obvious that for these large gate times of the experiment, the quality of the **AND** function realization is rather low. Specifically, if we consider ALT and LDH as Inputs 1 and 2, respectively, than the output at the logic **1**,**0** is too large as compared to the outputs at **0**,**0** and **0**,**1**, and is not sufficiently separated from that at **1**,**1**, to safely discriminate these two input combinations.

The "Filter" section of Scheme 1 shows the added filtering process. It involves reduction of $NAD^+$ back to NADH, which has two competing effects. On one hand, part of the output is suppressed: "recycled" back into NADH. This reduces the output at *all* the logic points, although on the relative scale the effect is more profound at those with low output values (**0**,**1**, **0**,**0** and **1**,**0**). However, the fact that the concentration of NADH is not decreased as much as for the non-filter system, partly prevents significant suppression of the net gate output when the input enzyme levels are both high, at **1**,**1**, The latter property allows us to avoid too much loss of the overall signal resolution (which could be another source of increasing the relative noise levels). The resulting measured gate function response is shown in Figure 2. Noise levels will be addressed quantitatively later, and we will discuss a proper selection of the gate- and filter-machinery chemicals to have the **AND** gate in the proper regime of functioning.

Figures 1 and 2 illustrate the issues involved in the use of biochemical processes to mimic logic-gate functions. We consider the output concentration, $[NAD^+](t_g)$, at a convenient (for applications) "gate time," $t = t_g$, here 600 sec, as a function of the initial concentration of the two inputs, $[ALT](t = 0)$ and $[LDH](t = 0)$. Here and below, the arguments in parentheses denote the time, $t$. The output can also be controlled via its parametric dependence on the initial concentrations of the other "gate machinery" and "filter-process machinery" chemicals identified in Scheme 1. Detailed modeling of the chemical kinetics of a function of two or three coupled enzymatic processes is complicated and impractical with the quality of data typically available in mapping out biochemical-logic gate responses. Indeed, enzymatic kinetics in these systems is quite noisy, with levels of noise in various data at least of the order of several percent of the data spread ranges. However, as emphasized in earlier works,[14,32,42,47] we only require an approximate fitting procedure to map out the response surface and estimate various noise effects (to be mentioned shortly) *near the logic-point values of the inputs*, as well as dependence on control parameters, for a possible gate-function optimization. This approximate, semi-quantitative model will be developed in the next section.

The logic-point values for the inputs, given in the preceding section, were selected at the highest published "normal" ranges and lowest "pathophysiological" ranges. The logic-point values for the output are set by the gate-function itself, as further discussed in the concluding section. Our gate-function optimization was initially largely done by adjusting the experimental control (gate-machinery and filter-machinery) parameters to get the process functioning in a proper regime, as detailed in the next section. However, the concluding section on results and discussion, reports gate-function quality analysis based on the developed model. This is best done in terms of the logic-range variables defined as follows.

To map the gate-function response surface, we consider for instance the initial concentration of Input 1, [ALT](0), not only at it logic-point values, which are $[\text{ALT}]_0(0) = 0.02$ U/mL, $[\text{ALT}]_1(0) = 2$ U/mL, but also at values inside and somewhat outside this range. Here and below, the subscripts **0** and **1** refer to the logic-point values. However, the actual noise-property analysis is best carried out in terms of the rescaled variable *x*,

$$x = \frac{[\text{ALT}](0) - [\text{ALT}]_0(0)}{[\text{ALT}]_1(0) - [\text{ALT}]_0(0)} \tag{3}$$

$$y = \frac{[\text{LDH}](0) - [\text{LDH}]_0(0)}{[\text{LDH}]_1(0) - [\text{LDH}]_0(0)} \tag{4}$$

$$z = \frac{[\text{NAD}^+](t_g) - [\text{NAD}^+]_0(t_g)}{[\text{NAD}^+]_1(t_g) - [\text{NAD}^+]_0(t_g)} \tag{5}$$

where the added equations define similar rescaled variables for the second input (*y*) and for the output (*z*).

Figure 2, showing a well-defined **AND** function, then illustrates various possible sources of noise that impede networking of several gates. The most obvious is the noise in the data and the uncertainties in defining the logic values. The noise in the data should be minimized relative to the spread of the logic values, such as, for the output, $[\text{NAD}^+]_1(t_g) - [\text{NAD}^+]_0(t_g)$. The gate-function itself can generate such noise both inherently and due to the imprecise realization of the logic-output values as compared to the convenient values desirable for networking the gate to feed its output into the next information/signal processing step. Here the residual spread between the three "**0**" outputs is the most obvious. In addition, the gate itself can actually amplify noise in the input. Avoiding this has been the main reason in the need for filtering to achieve "sigmoid" response, and the primary motivation for considering the whole gate-response surface outside the immediate vicinity of the four logic values. For smooth response surfaces and for signals which are not too widely spread, the noise amplification factor (if it is larger than 1) or suppression factor (if smaller than 1), can be estimated simply by calculating $|\vec{\nabla}z(x,y)|$, i.e., the slope of the gate-function response surface in terms of the rescaled variables, Equations (3-5). All this is addressed quantitatively in the concluding section.

**KINETIC MODELING OF THE GATE FUNCTION WITH ADDED FILTER**

In this section, we outline an approximate kinetic model which suffices for a semi-quantitative analysis of the **AND** gate functioning with the added filter process, as sketched in Scheme 1. The enzymatic cascade is initiated by the function of the enzyme ALT, the kinetics of which[48-51] can be to a good approximation described by the reactions[48]

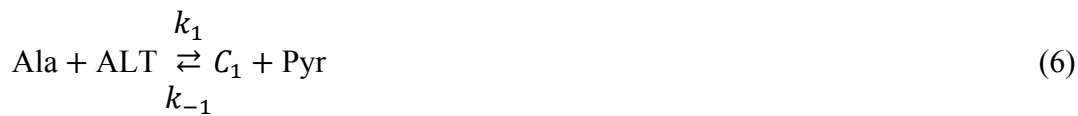

$$\text{Ala} + \text{ALT} \underset{k_{-1}}{\overset{k_1}{\rightleftarrows}} C_1 + \text{Pyr} \tag{6}$$

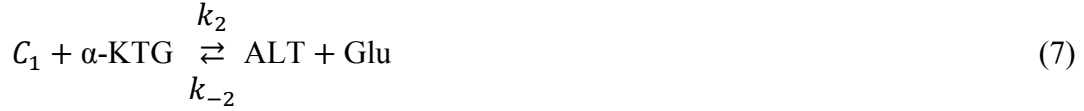

$$C_1 + \alpha\text{-KTG} \underset{k_{-2}}{\overset{k_2}{\rightleftarrows}} \text{ALT} + \text{Glu} \tag{7}$$

where $C_1$ denotes a complex. We chose the conditions with a large supply of Ala, [Ala](0) = 200 mM, in order to have fast reaction steps and thus plentiful feed of Pyr into the LDH part of the cascade. Therefore, for our modeling, the back-reactions (rate constant $k_{-1,-2}$) in Equation (6-7) were neglected, and we can also ignore the depletion of Ala (set it to its initial value for all times considered). Similarly, we also took a large initial supply [α-KTG](0) = 10 mM.

Such a selective adjustment of relative rates of processes involved, de facto constitutes an experimental "optimization" of the gate-function system to yield a high-quality **AND**-gate realization. Additional adjustments will be described shortly, and some of these rely on the presence of the filter part of the cascade. Thus the assumptions leading to our few-parameter model are not precise without the filter, as commented on below and seen for low LDH concentrations in Figure 1.

In order to minimize the number of rate constants used in our model, and given the observations on the fast-reaction nature of the processes involved in the ALT part of the cascade, we can assume that this sub-system is practically always in the steady state, and furthermore, the fraction of the enzyme in the complex is approximately proportional to the initial enzyme concentration. Therefore, in the realized regime we can approximately parameterize the whole process biocatalyzed by ALT, by a single parameter $R_{\text{ALT}}$, such that the effective process and its rate constant are

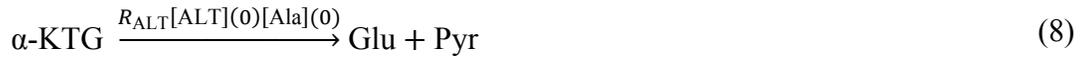

$$\alpha\text{-KTG} \xrightarrow{R_{\text{ALT}}[\text{ALT}](0)[\text{Ala}](0)} \text{Glu} + \text{Pyr} \tag{8}$$

Thus, we assume that the effective irreversible process rate constant is approximately linear in the input [ALT](0) and also in the initial concentration of [Ala](0). The overall rate constant is then proportional to the product of these quantities, times the additional adjustable parameter $R_{\text{ALT}}$. In fact, the concentration of α-KTG is also expected not be much depleted. However, we prefer to keep it a time-dependent quantity because we later reference modeling for increased initial values of Glc6P on the "filter" side of the cascade which can result in a non-negligible consumption of α-KTG (verified by considering specific concentration values, not detailed here). Also, this does not add parameters to the model.

The output of the previously considered cascade step, initiates the next, LDH-biocatalyzed, part of the system,[52-55]

$$\text{NADH} + \text{LDH} \underset{k_{-3}}{\overset{k_3}{\rightleftarrows}} C_2 \tag{9}$$

$$C_2 + \text{Pyr} \underset{k_{-4}}{\overset{k_4}{\rightleftarrows}} C_3 + \text{Lac} \tag{10}$$

$$C_3 \underset{k_{-5}}{\overset{k_5}{\rightleftarrows}} \text{LDH} + \text{NAD}^+ \tag{11}$$

where $C_{2,3}$ are complexes. Again, for a schematic description with as few parameters as possible, we assume that a single-rate, irreversible process can be used in the realized regime,

$$\text{NADH} + \text{Pyr} \xrightarrow{R_{\text{LDH}}[\text{LDH}](0)} \text{Lac} + \text{NAD}^+ \tag{12}$$

Similar to Equation (8), here again a single adjustable parameter, $R_{\text{LDH}}$, is used in the rate constant, multiplied by the initial input [LDH](0). Indeed, here the reaction steps, Equations (9-11), are fast and practically irreversible, driven by large supply of NADH and Pyr. The last equation will be rate-determining, and here also, in the steady-state regime the fraction of the enzyme in the complexes will be more or less constant. The effective description, Equation (12), can then be used. However, when a significant fraction of NADH is used up *and* the initial supply of LDH (Input 2) is low, then these assumptions break down (the first reaction slows down). Thus, the model becomes inaccurate for small [LDH](0) values, especially when there is no "filter" which back-converts NAD$^+$ to NADH. This can be seen in Figure 1.

The "filter" part of the cascade is biocatalyzed by G6PDH, the mechanism of action of which is not fully sorted out and likely can follow several pathways.[56-60] We chose the pathway reported to have the highest rate,[56] described by the reactions,

$$\text{Glc6P} + \text{G6PDH} \underset{k_{-6}}{\overset{k_6}{\rightleftarrows}} C_4 \tag{13}$$

$$C_4 + \text{NAD}^+ \xrightarrow{k_7} \text{G6PDH} + \text{6-PGluc} + \text{NADH} \tag{14}$$

and we also ignore possible weak allosteric effects (homotropicity of Glc6P) for this enzyme. Here $C_4$ is a complex. The single-parameter effective description for this part of the cascade was assumed to be

$$\text{NAD}^+ + \text{Glc6P} \xrightarrow{R_{\text{G6PDH}}[\text{G6PDH}](0)} \text{6-PGluc} + \text{NADH} \tag{15}$$

The introduction of the adjustable parameter, $R_{\text{G6PDH}}$, here is similar to that in Equations (8) and (12). This is based on that a large supply of Glc6P was used to drive the reaction fast and irreversible. As before, we assume that the fraction of the enzyme in the complex is approximately fixed for fast reactions in the steady state (and thus absorbed in the definition of the parameter $R_{\text{G6PDH}}$).

The model now reduces to rate equations for the time-dependence of concentrations of all those chemicals which are not assumed approximately constant. We only show one illustrative equation here to clarify the notation in Equations (8,12,15), and to indicate the units used. For example, the rate equation for NADH is

$$\frac{d[\text{NADH}](t)}{dt} = -R_{\text{LDH}}[\text{LDH}](0)[\text{NADH}](t)[\text{Pyr}](t) \\ + R_{\text{G6PDH}}[\text{G6PDH}](0)[\text{NAD}^+](t)[\text{Glc6P}](t) \tag{16}$$

The first two factors in each term in the overall rate of variation of [NADH]($t$), i.e., on the right-hand side in Equation (16), represent the effective rate constants marked above the arrows in our schematic Equations (8,12). Concentrations of all the chemicals were either initially given in mM, or, for the three enzymes, expressed in these units by using the conversion factors $7.39 \cdot 10^{-5}$ mM/(U·mL$^{-1}$), $0.94 \cdot 10^{-5}$ mM/(U·mL$^{-1}$), and $1.18 \cdot 10^{-5}$ mM/(U·mL$^{-1}$), for ALT, LDH, and G6PDH, respectively. The conversion factors were estimated based on the enzyme activity provided by the supplier (Sigma-Aldrich) and known molecular mass values.[61] As a result, the model fitting yields the rate parameters $R_{LDH}$ and $R_{G6PDH}$ in units of 1/(mM$^2$·sec), with the same units for $R_{ALT}$ which was defined in Equation (8).

**RESULTS AND DISCUSSION**

Since the model was devised to work in the regime of "filtered," well-defined **AND**-gate functioning of the system, we used not just the gate-time (600 sec) experimental data, but actually the full set of the time-dependent output signals, [NAD$^+$]($t$), of the experiments with the filtering process on, for fitting the rate parameters values. These were measured for all the 108 (= 12×9) input combinations listed in the experimental section, in time increments of 1 sec. The fitted parameter values are as follows: $R_{ALT} = 2.1 \cdot 10^{-4}$ mM$^{-2}$ sec$^{-1}$, $R_{LDH} = 3.8 \cdot 10^{-1}$ mM$^{-2}$ sec$^{-1}$, $R_{G6PDH} = 9.5 \cdot 10^{-3}$ mM$^{-2}$ sec$^{-1}$. These parameters were used to calculate the model surfaces shown in Figures 1 and 2.

Generally, the present model is approximate, but it provides a good-quality semi-quantitative fit for the measured data not only with the filter (Figure 2), but also without the filter (with the same rate parameter values: Figure 1), except, as anticipated, for low input concentrations of LDH in the latter case. Our focus here is on results at the gate time. However, in Figure 3 we also illustrate how do model-calculated curves approximate the actual time-dependent data for several input values. The latter were selected to have typical examples of the time-dependent variation of the data sets measured for the "filtered" case, as well as a comparison with the non-filtered case for the logic **1,0** inputs (for which the filtering effect is maximal, cf. Figures 1 and 2).

Next, we concentrate on the vicinity of the four logic-point input combinations, and outputs measured at the gate time, aiming at using the model-fit results to evaluate "quality" measures (to be introduced shortly) of the realized **AND**-gate. As mentioned in the introduction, the logic-**0** and **1** values of the two inputs are pre-set by the actual expected application of the gate system. However, the logic-**0** and **1** values of the output, are determined by the gate function itself and therefore actually depend on the selection of the controllable "machinery" parameters (chemical concentrations), as well as, in principle, on the chemical and physical parameters of the solution environment in which the gate system is operated. It is natural to set the output logic-**1** value for the **AND** function, $[\text{NAD}^+]_1(t_g)$, as that measured for **1**,**1** inputs. Because of the noise in the data, this can be selected as a convenient "round" value near the measured data for the **1**,**1** input combination. Another option, that we used here, is to utilize the value predicted by the model (this works as long as the model is sufficiently accurate). The selection of the logic-**0** for the output is complicated by that a certain part of the "noise" in the gate-function realization is the spread of the outputs at inputs **0**,**0**, **0**,**1** and **1**,**0**. For the "filtered" system they are not spread much (Figure 2), and we could use, for instance, their mean value. However, for the non-filtered system the realization of the **AND** function is of a low quality (Figure 1), and therefore for definiteness we preferred to define the logic-**0** of the output, $[\text{NAD}^+]_0(t_g)$, as the value at the **0**,**0** input. For example, for the parameters used for our "filtered" experiment, see Figure 2, we used the model-calculated values $[\text{NAD}^+]_0(t_g) \simeq 0.2 \cdot 10^{-2}$ mM and $[\text{NAD}^+]_1(t_g) \simeq 29.8 \cdot 10^{-2}$ mM.

Noise in the gate output, clearly seen in the experimental data in Figures 1 and 2, is determined by several factors. First, the gate function itself can have certain degree of noise in its actual realization, it means the output values are not always precise but can be distributed. Second, there is also a systematic spread/drift of values, which is best exemplified here by the different outputs at the three logic inputs that are expected to yield logic-**0** output. Another contribution to the noise in the output is due to the noise in the actual inputs, which is "transmitted" via the gate function and can be amplified or suppressed. While a more careful discussion in terms of network designs is important for larger networks,[42,47] we note that, at the level of considering an optimal realization for each gate as a possible *network element*, we try to minimize noise amplification (maximize its suppression) by ensuring that the slopes of the response surfaces are as small as possible, hopefully, less than 1 at all the logic points. This criterion[32] is suitable only for smooth gate function surfaces, which is the case here. In fact, the primary reason to adding the "filter" processes has been to reduce the slopes at all the logic points to well below 1 (as will be quantified later). In our case, for large gate times, the added "filter" actually enables the gate to carry out a relatively accurate **AND** function (compare Figures 1 and 2).

Our optimal selection of the gate-function (with filter) parameters, was originally based on fine-tuning during the preliminary trial-and-error experiments. In order to substantiate the result and also quantify it within the proposed model, let us consider some criteria for gate functioning. We compare the no-filter case with the filtered systems for the original and modified control-parameter values. For the latter, we consider the cases of halving and doubling the "filter machinery" chemical Glc6P initial concentration: [Glc6P](0). Indeed, here it was found to have a significant effect on the gate functioning, whereas earlier studies[32] suggest that generally the parameters controlling the "gate" part of the process (see Scheme 1) have to be changed by orders of magnitude to have an appreciable effect. Interestingly, this is another useful feature of the recent discovery[34,35] of that simple added chemical processes can convert convex biochemical response to sigmoid. Our present work is the first such system demonstrating a "fully sigmoid" two-input gate, and furthermore, one with the non-zero-concentration logic-**0** values. It transpires that the "filter machinery" chemicals generally have a more profound effect on the overall process than the "gate-part machinery" ones, when changed by moderate factors.

For the filtered system with the originally selected experimentally-optimized parameters, the slope $|\vec{\nabla}z(x,y)|$ in terms of the rescaled "logic-range" variables, Equations (3-5), is mapped out as a function of the original (not scaled) inputs in Figure 4. As expected for a good-quality filter, at the logic-point values the slopes are well below 1, resulting in noise suppression. Table 1 (see page 21) lists the estimates (in percent of the logic-range intervals between **0** and **1**) of the noise possible in the input that will not cause their deviations past the slope-1 lines marked in the figure. This "noise tolerance" percentage is given as the smallest of the four values at the logic points. Similar analysis was carried out for the modified values of [Glc6P](0), see Table 1. For the non-filtered system, the noise tolerance listed is zero, because even without developing a

precise model for this case, it is clear from Figure 1 that the largest slope of the gate-response function is significantly larger than 1. This is typical for biocatalytic reactions.[14,32,42] Thus, in addition to its built-in inaccuracy, the non-filtered **AND** gate will also significantly amplifies input noise as it is passed to the output (numerical estimates not detailed here suggest the value of $|\vec{\nabla} z(x,y)| \simeq 4.5$, indicating input-to-output noise amplification by a factor of approximately 450%).

The estimates reported thus far, generally confirm our expectation that, increasing the filtering effect, here by introducing more Glc6P, generally improves handling of noisy input, by improving the "tolerance" as presented in Table 1. However, there is a tradeoff in that, the realized filtering mechanism generally decreases the overall signal intensity measured as the difference between the logic-**1** and **0** output values, see Table 1. Loss of intensity makes the system *relatively* more sensitive to the absolute value of the noise. We already discussed the fact that, there is a certain inherent noise generation by the gate function itself. For the present system the leading source of the "built-in" noisiness is simply the spread between the output values at the inputs that are expected to yield logic-**0** outputs. This quantity is also estimated as percentage of the signal intensity in Table 1. All the values in Table 1 were rounded to whole percents for clarity, and all but one were based on model estimates. Since for the *non-filter* case the model is not accurate near input $y = 0$ (low LDH), the model-based "spread" estimate would be particularly inaccurate. The actual value presented in Table 1, was thus, for this single entry only, estimated directly from the non-filter experimental data (Figure 1).

Values presented in Table 1, suggest that the selected initial Glc6P concentration offers a good balance between the tolerance to noise in the input, the inherent noisiness of the **AND** gate realization, and at the same time not too much loss in the overall signal intensity. Specifically, as a network element, our gate generates less spread in the data than what it could accept from other network elements, and this is accomplished without too much loss of intensity.

In summary, our present study has offered the first experimental as well as model-substantiated demonstration that, the two-input sigmoid response (double-sigmoid) can be achieved in biomolecular logic by adding a "filter" process. The filtering mechanism was found to be useful at physiological input concentrations, including non-zero logic-**0** values. Furthermore, the filtering approach considered here—an added chemical reaction involving the *output*[34-36]—is advantageous as compared to approaches involving one[33] or both of the *input* chemicals, because it offers a more straightforward approach to achieve the double-sigmoid response. We have also learned that, to control the **AND**-gate performance, the "filter machinery" process parameters require much less substantial adjustments than order-of-magnitude changes needed for controlling the gate properties via the "gate machinery" parameters.

## Acknowledgements

Research funding by the NSF, via awards CCF-1015983 and CBET-1066397, is gratefully acknowledged.

**Table 1:** Gate-quality measures for several initial values of the "gate machinery" chemical Glc6P. "Noise tolerance" evaluates the percentage on input noise that the gate can tolerate without amplifying it. "Signal intensity loss" indicates the decrease in the signal range (in the difference between the logic-**1** and **0** output values) as percentage of the non-filtered signal range. "Built-in noise" is estimated by considering the spread of the outputs at inputs **0**,**1**, **1**,**0** from the selected logic-**0** output level (at **0**,**0** input), as percentage of the overall signal intensity.

| System | [Glc6P](0) (mM) | Noise tolerance | Signal intensity loss | Built-in noise |
|:---:|:---:|:---:|:---:|:---:|
| no filter | 0 | 0 | 0% | 29% |
| realized filter | 1.2 | 8% | 12% | 5% |
| model | 0.6 | 3% | 4% | 7% |
| model | 2.4 | 18% | 46% | 3% |

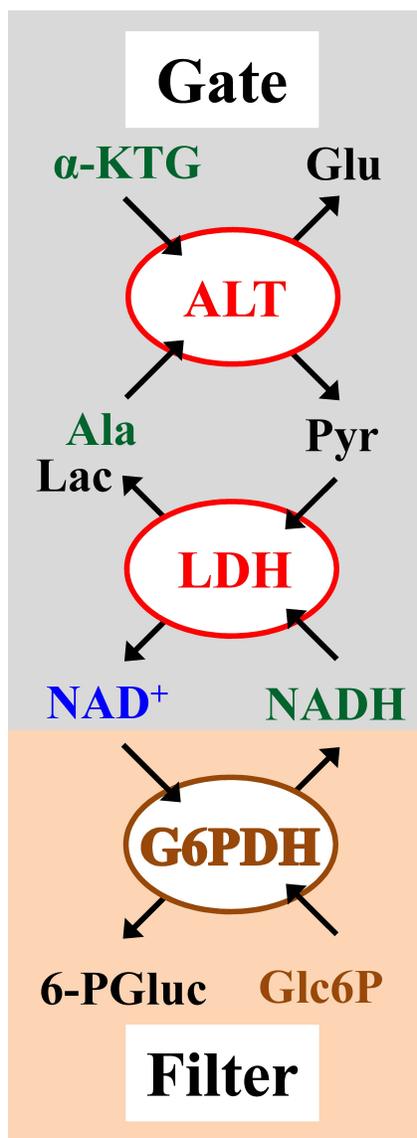

**Scheme 1.** Biocatalytic cascade realizing the AND-gate function with the filter process added. The following abbreviations for products and intermediates are used: Glu for glutamate, Pyr for pyruvate, Lac for lactate, and 6-PGluc for 6-phospho-gluconic acid. All the other notations are defined in the text. Color coding: green – "gate machinery" chemicals; black – reaction intermediates and byproducts; red – enzyme inputs; brown – "filter machinery" chemicals; blue – output chemical.

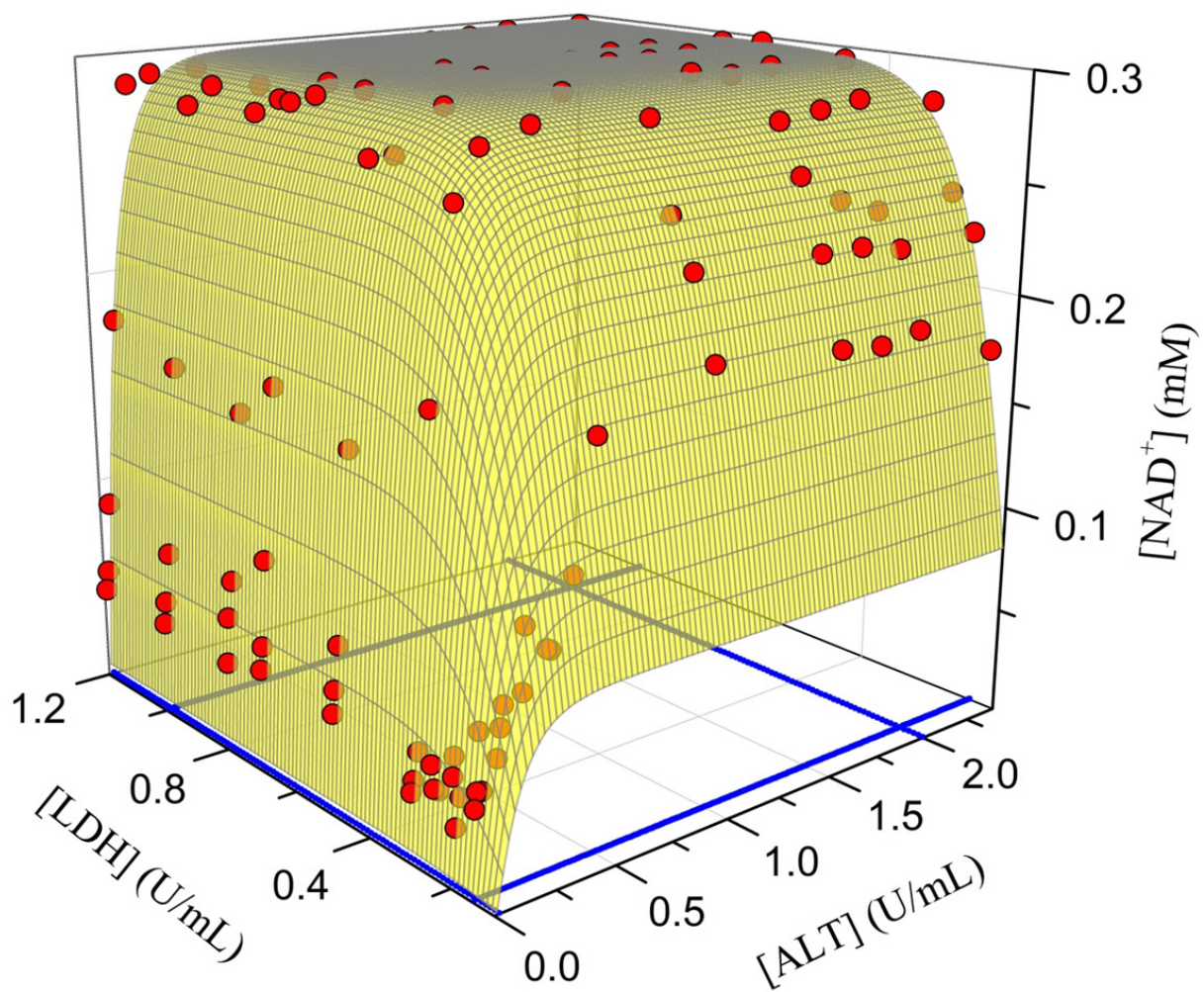

**Figure 1.** Spherical symbols show the 108 (12×9) experimental data points for the system without the "filter" process. These were measured at the gate time $t = 600$ sec, for various values of the two input enzyme concentrations as detailed in the text. The blue lines are drawn at the selected logic-**0** and **1** values of the inputs and thus delineate the "logic" range for mapping out the gate-response function. The surface was calculated from the model described in the text, with the parameters fitted based on the full time-dependent data set for the "filtered" system.

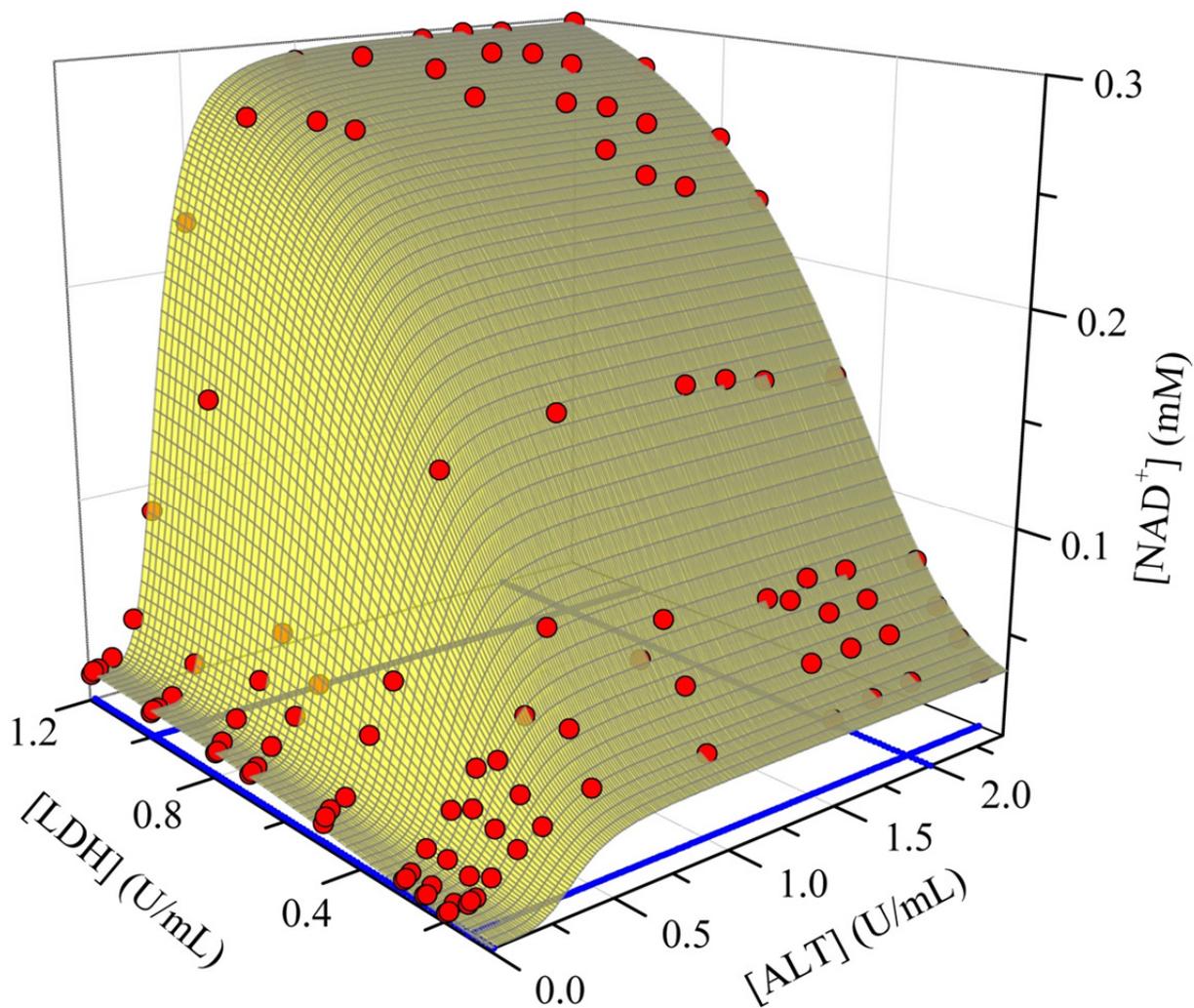

**Figure 2.** Spherical symbols show the 108 (12×9) experimental data points for the system with the "filter" process, measured at the gate time $t = 600$ sec. The blue lines mark the selected logic-**0** and **1** values of the inputs. The surface was calculated from the model described in the text, with the parameters fitted based on the full time-dependent data set for the "filtered" system.

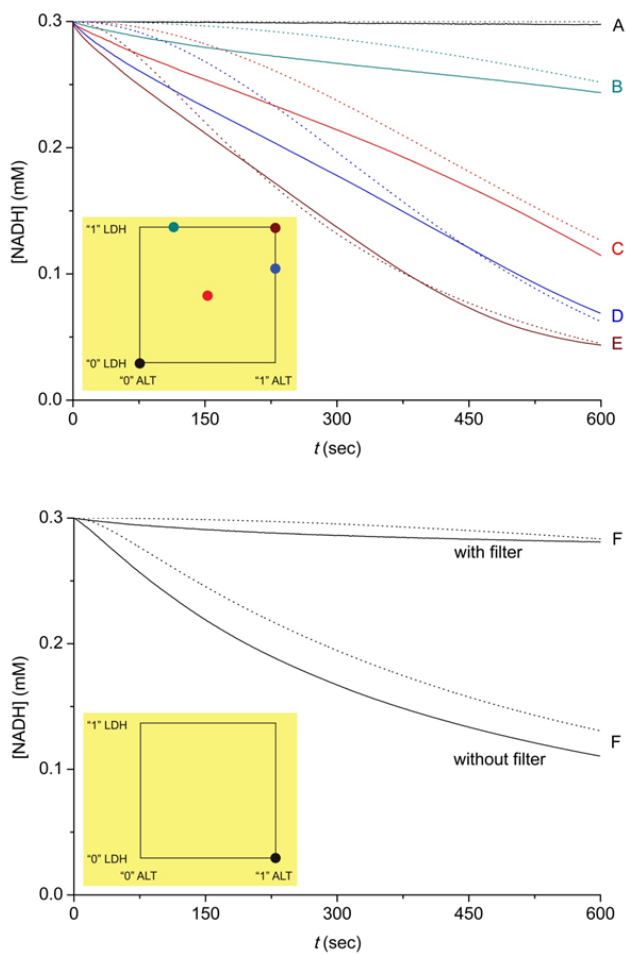

**Figure 3.** Illustration of the semi-quantitative nature of the few-parameter model developed for the "filtered" system. The experimental data, measured in steps of 1 sec, are drawn as solid lines. The model results are shown as dotted lines. Out of the total 108 time-dependent data sets measured, five are shown in the top panel, selected such that the curves do not much overlap. The degree of agreement between the model and these data sets is similar for other data. The bottom panel compares the data and its model fits for the "filtered" and "non-filtered" systems, for the case of the logic **1,0** inputs, for which the filtering effect is the most significant, as seen in Figures 1 and 2. The data sets presented, were measured for input combinations as color-coded in the insets and corresponding to the following logic-range values, A: (0,0), B: (1/4,1), C: (1/2,1/2), D: (1,2/3), E: (1,1), F: (1,0).

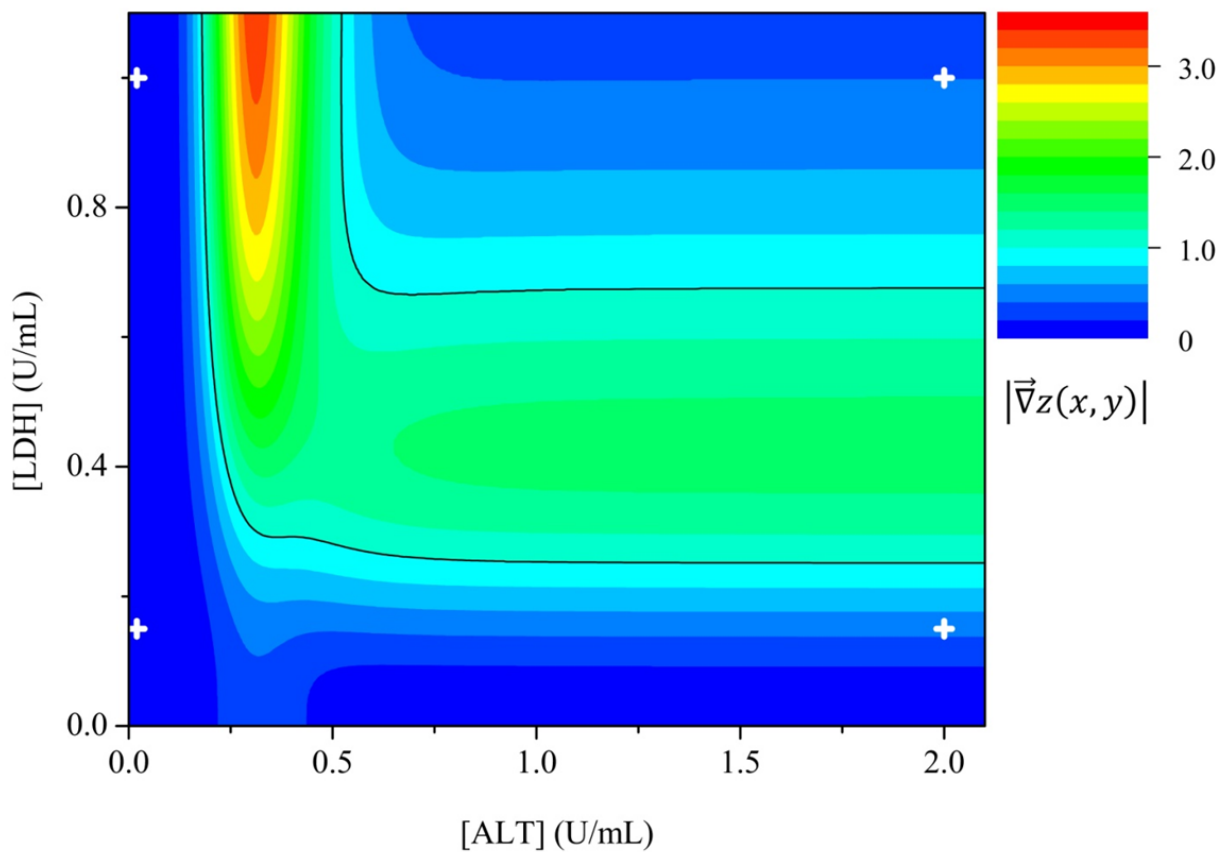

**Figure 4.** Color-coded contour plot of the absolute value of the gradient vector in terms of the rescaled variables, $|\vec{\nabla}z(x,y)|$, calculated from the fitted model for the experimentally realized "filtered" system parameters. The solid black lines indicate the contours of $|\vec{\nabla}z(x,y)| = 1$. The four white crosses mark the logic inputs **0,0**, **1,0**, **0,1**, **1,1** in the plane of the input enzyme concentrations.